\documentclass[useAMS,usenatbib]{mn2e}
\usepackage{graphicx,natbib,times,mathptm,bm,amssymb}
\usepackage[fleqn]{amsmath}

{\newif\ifnotend
\notendtrue
\def\veclist{ABCDEFGHIJKLMNOPQRSTUVWXYZabcdefghijklmnopqrstuvwxyz.}
\def\top#1#2.{#1}
\def\tail#1#2.{#2.}
\loop\expandafter\xdef\csname v\expandafter\top\veclist\endcsname%
{{\noexpand\bm\expandafter\top\veclist}}
\edef\veclist{\expandafter\tail\veclist}
\if\veclist.\notendfalse\fi\ifnotend\repeat}
\def\d{{\rm d}}
\def\e{{\rm e}}
\def\i{{\rm i}}
\def\pr{{\mathop{\hbox{p}}}}

\def\setofA{\bm{A}}
\def\setofs{\bm{s}}

\def\setofyhat{\tilde{\bm{y}}}
\def\setofl{\bm{l}}
\def\setoflhat{\tilde{\bm{l}}}
\def\setofb{\bm{b}}
\def\setofbhat{\tilde{\bm{b}}}

\newcommand{\appropto}{\mathrel{\vcenter{
  \offinterlineskip\halign{\hfil$##$\cr
    \propto\cr\noalign{\kern2pt}\sim\cr\noalign{\kern-2pt}}}}}

\bibpunct{(}{)}{;}{a}{}{,}

\def\macromodel{\Theta}
\def\micromodel{\zeta}
\def\galaxymodel{\beta}
\def\transfermodel{\alpha}
\def\macroasumps{\eta}

\def\cov{\mathop{\hbox{Cov}}}

\begin{document}

\bibliographystyle{mn2e}

\title[3d extinction mapping using GRFs]{Three-dimensional extinction mapping using Gaussian random fields}

\author[Sale and Magorrian]{S.~E.~Sale and  J. Magorrian\\
Rudolf Peierls Centre for Theoretical Physics, Keble Road, Oxford OX1 3NP, UK\\}

\date{Received .........., Accepted...........}

\maketitle

\begin{abstract}
  We present a scheme for using stellar catalogues to map the three-dimensional distributions of extinction and dust within our Galaxy.
  Extinction is modelled as a Gaussian random field, whose covariance function is set by a simple physical model of the ISM that assumes a Kolmogorov-like power spectrum of turbulent fluctuations.
  As extinction is modelled as a random field, the spatial resolution of the resulting maps is set naturally by the data available; there is no need to impose any spatial binning.
  We verify the validity of our scheme by testing it on simulated extinction fields and show that its precision is significantly improved over previous dust-mapping efforts.
  The approach we describe here can make use of any photometric, spectroscopic or astrometric data; it is not limited to any particular survey.
  Consequently, it can be applied to a wide range of data from both existing and future surveys.

\end{abstract}

\section{Introduction}\label{sec:intro}

The extinction of starlight by interstellar dust is a nuisance for many astronomers.
It is particularly troublesome in our own Galaxy: our understanding of the Galactic disc is hampered by in-plane dust, a frustrating situation given that the disc contains the bulk of the Galaxy's stellar mass and is where the most interesting dynamical phenomena occur.
There are already many stellar surveys that contain a wealth of still-undigested data about the structure and dynamics of the Galaxy, including those covering the whole sky (2MASS), targeted at the Galactic Plane (e.g. IPHAS \citealt{Drew_Greimel.2005}) or looking out of the plane (e.g. RAVE, SDSS).
Many more, such as Gaia and Pan-STARRS and VPHAS+ \citep{Drew_Gonzalez-Solares.2014}, are underway.
Any attempt to make sense of the catalogues produced by these surveys inevitably involves constructing model Galaxies and comparing the predictions from these models to the observed catalogues.
For such a comparison to be meaningful, it is essential that the models' predictions take account of the extinguishing effects of the Galaxy's three-dimensional distribution of dust.

The well-known dust maps of, e.g., \citet{Schlegel_Finkbeiner.1998} or the \citet{PlanckCollaboration_Abergel.2013a} are not sufficient for this purpose, because they provide only the column density of dust integrated along each line of sight and have nothing to say about how the dust is distributed with distance.
Nevertheless, these two-dimensional maps do provide important boundary conditions that any three-dimensional map must satisfy.
There has been an increased focus on 3d mapping over the last several years.
\cite{Marshall_Robin.2006} compared 2MASS photometry to the Besan\c{c}on Galactic model to produce a map of the extinction in much of the Galactic plane. \cite{Majewski_Zasowski.2011} presented the Rayleigh--Jeans colour excess method, which relies on the use of infrared data.
\cite{Berry_Ivezic.2012} estimated the distance and extinction to stars from SDSS and 2MASS, treating each star individually, before then considering the mean extinction of stars in spatial bins.
\cite{Hanson_Bailer-Jones.2014}, building upon \cite{Bailer-Jones_only.2011}, proceeded along similar lines, though with a more sophisticated Bayesian method to estimate the distance and extinction to each star.
A method which employs hierarchical Bayesian models to simultaneously estimate the reddening--distance relationship along a line of sight, and the properties of the stars that trace it, was presented in \cite{Sale_only.2012}, superseding an earlier method by \cite{Sale_Drew.2009}.
This has been applied by \cite{Sale_Drew.2014} to construct a 3D extinction map of the Northern Galactic plane from the IPHAS DR2 catalogue \citep{Barentsen_Farnhill.2014}.
\cite{Green_Schlafly.2014} adopted a similar approach, though the algorithmic details differ.

None of these methods properly account for the small-scale, fractal structure of the ISM; most assume that the dust column does not vary on scales of $\sim 100\,\rm pc$.
This assumption is fundamentally flawed, however: imaging studies show that the dust distribution has structure on sub-parsec scales \citep[e.g. Figure 1 of][]{Barentsen_Vink.2011}, with interstellar scintillation work demonstrating that the ISM has structure on scales as small as $\sim$100~km \citep[e.g.][]{Spangler_Gwinn.1990}.
Ignoring this substructure not only results in a loss of detail, but also leads to a potential bias in the resulting dust maps.
\cite{Sale_only.2012} included a simple treatment for variations of extinction on small spatial scales, but \cite{Sale_Drew.2014} found that the limitations of this treatment produced the dominant contribution to the uncertainty in the extinction to a given position, which can be significant at shorter distances.

An additional fact that is largely ignored by the preceding methods is that neighbouring patches of the ISM are not independent; turbulent processes fed by a variety of energy sources \citep{Elmegreen_Scalo.2004} mix the gaseous components of the ISM.
As dust is a passenger in the gas flows it too is mixed, causing its density to be correlated over the scales on which turbulence operates.
Consequently, it is possible to constrain more tightly the density of components of the ISM by building these expected large-scale correlations into the mapping algorithm.
This was recognised by \cite{Vergely_FreireFerrero.2001} who applied geophysical mapping techniques from \cite{Tarantola_Valette.1982} to mapping the interstellar NaI and HII density.
Subsequently, \cite{Vergely_Valette.2010} and \cite{Lallement_Vergely.2014} have applied the same approach to mapping extinction in the local ISM.
A similar approach was used by \cite{Pichon_vergely.2001} to map the Lyman-$\alpha$ forest.

In this paper we present a method for constructing three-dimensional extinction maps based on modelling the dust density as a random field.
Our method allows extinction to vary on all spatial scales by building in a simple physical model of the turbulence that produces structure in the ISM.
The paper is organised as follows.
Section~\ref{sec:dustmodel} describes our model for the spatial distribution of dust.
In Section~\ref{sec:mapmaking} we tackle the general dust-mapping problem using an extension of the hierarchical Bayesian approach introduced by \citet{Sale_only.2012}.
We identify the posterior probability distributions that are most important for applications, and explain how they can be constructed from observations.
To illustrate the application of the method, in Section~\ref{sec:2d} we use a simple two-dimensional dust-mapping problem.
Section~\ref{sec:discussion} compares our scheme to other methods and in section~\ref{sec:summary} we identify future work.

\section{Model for the spatial distribution of dust}\label{sec:dustmodel}

\subsection{Motivation}

No region of the ISM lives in isolation: the properties of the ISM at a given location are correlated with those of other positions, both in the immediate locality and at greater distances.
These correlations are a product of the turbulent processes which shape the ISM over a range of spatial scales.
We would like to establish a statistical framework to describe the distributions of dust and extinction, including a physical treatment of the correlations that exist as a product of turbulence.

\subsubsection{Spatial scaling}\label{sec:spatial_scaling}

\cite{Kolmogorov_only.1941} proposed a now classic theory of turbulence, whereby energy is introduced to a medium at large scales. 
These processes cause the formation of large structures. Subsequently energy ``cascades'' down to smaller length scales, across an inertial range, producing structures of a corresponding size as it flows. 
Eventually, at small scales, viscosity becomes dominant and the energy is dissipated. 
In this picture dust is essentially a passenger, being dragged along by the gaseous components of the ISM which are themselves experiencing turbulence. 
However, as such it acts as a tracer of the gas distribution and so exhibits a similar density structure on the scales which are governed by turbulence.

One feature of Kolmogorov turbulence is that the density power spectrum of the medium takes the power-law form 
\begin{align}
|\tilde\rho(\vk)|^2 \propto |\vk|^{-\gamma}, 
\label{eq:kolmopower}
\end{align}
where $\vk$ represents the (three-dimensional) wavenumber and the exponent $\gamma=11/3$ across the inertial range \citep[e.g.,][]{Armstrong_Rickett.1995}. 
Observations confirm that this form is valid for the electron density over a wide range of scales in the diffuse ISM \citep{Armstrong_Rickett.1995, Chepurnov_Lazarian.2010}.
Similarly,
HI surveys \citep[e.g.,][]{Lazarian_Pogosyan.2000} and MHD simulations also obtain power-law power spectra, although
these often display different slopes \citep[e.g.,][]{Dickey_McClure-Griffiths.2001,Padoan_Jimenez.2004};
the inclusion of magnetic fields (both in reality and MHD simulations) and the fact that the real ISM is not incompressible breaks a number of the assumptions made by Kolmogorov, giving rise to this discrepancy.
Therefore, in the following we describe turbulence as being ``Kolmogorov-like'' if the power spectrum of its density fluctuations takes the power-law form~\eqref{eq:kolmopower}, even if the power-law index $\gamma\ne11/3$.

\cite{Lazarian_Pogosyan.2000}, amongst others, demonstrate that the power spectra of projections of the density field, which include extinction, should follow a broken power-law. 
Specifically, the power spectrum slope changes at a scale corresponding to the transition between thin and thick screens of ISM. 
Maps of integrated Galactic dust emission show power-law angular power spectra \citep{Schlegel_Finkbeiner.1998, Kiss_Abraham.2003, Roy_Ade.2010} and similar spectra are also retrieved from 2D maps of extinction \citep{Melbourne_Guhathakurta.2004, Brunt_only.2010}.

\subsubsection{Distribution of density fluctuations at fixed position}
Both simulations \citep[e.g.,][]{Ostriker_Stone.2001} and analytical arguments \citep[e.g.,][]{Nordlund_Padoan.1999} indicate that, under a wide variety of conditions, the log density $\log\rho(\vr)$ is approximately normally distributed, with a variance that depends on the local Mach number.  
But, note that \citet{Hopkins_only.2013} suggests an alternative distribution that may provide a better fit.

However, observations do not provide a direct probe of dust density, and so, for this paper, we are not immediately concerned with its probability distribution.
Instead we view projections of the density field, such as extinction, which is itself directly related to the column density.
For an observer located at the origin, the monochromatic
\footnote{Following \cite{McCall_only.2004}, \cite{Sale_Drew.2009} and \cite{Bailer-Jones_only.2011}, we consider only monochromatic extinctions in this paper, as they depend only on the dust column and not the properties of the observed star.}
 extinction to a point at distance~$s$ having galactic coordinates $(l,b)$ is 
\begin{equation}
  A(l,b,s)\equiv \int_0^s
   \kappa\rho(l,b,s')\,\d s',
\label{eq:Adefn}
\end{equation}
where $\kappa$ is the opacity and, for clarity, we have suppressed the
wavelength dependence of both $A$ and~$\kappa$ and any spatial dependence of~$\kappa$.
Clearly, the relationship between the statistical properties of $A$
are easy to determine from those of $\rho$.  The relationship
between the properties of $\log\rho$ and $\log A$ is not so
straightforward, however.

Nevertheless, \cite{Ostriker_Stone.2001} reason that $\log A$ should have an approximately Gaussian distribution.
Their argument is as follows.
Assume that the ISM along each line of sight is subject to many independent regions of compression or rarefaction. Each such region increases/decreases the density of the ISM by some fraction in the part of the sightline affected. Therefore it affects the projected density, along the entire line of sight, by another fraction, which is much closer to one.
Consequently, the logarithms of these fractions are small.
As the extinction contrast is the product of these fractions, it follows that the $\log$ extinction contrast is the sum of the logarithms.
Given that the compression or rarefaction events are assumed to be independent, it follows that the $\log$ extinction contrast should follow a Gaussian distribution and so $\log A$ should be normally distributed.
This is borne out by both observations \citep{Lombardi_Alves.2006,Kainulainen_Beuther.2009,Froebrich_Rowles.2010,Alves_Lombardi.2014} and simulations \citep{Ostriker_Stone.2001,Vazquez-Semadeni_Garcia.2001}, with only a small departure at the highest extinctions that occurs in regions of strong star formation.

\subsection{A model for the statistical properties of the extinction distribution}

The fundamental assumption of our method is that the logarithm of the extinction can be modelled as a semi-stationary Gaussian random field.
To set the statistical properties of this field we use a model for the dust density that is based on the considerations above.
Before describing our model in detail we first recall some basic properties of random fields in general, and Gaussian and semi-stationary random fields in particular.

\subsubsection{Gaussian random fields}\label{sec:grf_intro}
By a random field~$f(\vr)$ we mean a random process defined on three-dimensional space.
We use $f$ to refer both to the random process itself and to a particular function drawn from the random process.
Given a set of $N$ points, $\vr_1,...,\vr_N$, there is then a well-defined joint probability density function $\pr(\vf)$ for the values $\vf\equiv (f(\vr_1),...,f(\vr_N))$ of $f$ at those points.
The first few moments of this pdf are a convenient (but incomplete) summary of the underlying random field.  The most important are the mean and covariance, defined via
\begin{equation}
  \begin{split}
  \bar f(\vr)&\equiv\langle f(\vr)\rangle,\\
\cov(f(\vr_1),f(\vr_2)&\equiv
\left\langle (f(\vr_1)-\bar f(\vr_1)(f(\vr_2-\bar f(\vr_2)\right\rangle\\
&=\left\langle f(\vr_1)f(\vr_2)\right\rangle -\bar f(\vr_1)\bar f(\vr_2),
  \end{split}
\end{equation}
where $\langle f(\vr)\rangle$ denotes the value of $f(\vr)$ averaged over many realizations of~$f$.  The quantity $\left\langle f(\vr_1)f(\vr_2)\right\rangle$ that appears in the expression for the covariance is the correlation function.

A random field $f$ is {\it stationary} if it is invariant under spatial translations.
Clearly the expectation value of a stationary field is independent of position, $\bar f(\vr)=\hbox{constant}$, and the covariance functions $\cov[f(\vr_i),f(\vr_j)]=\mathcal{C}(|\vr_1-\vr_2|)$ depends only on the scalar distance between points.
By the Wiener--Khinchin theorem, the power spectrum $|\tilde f(\vk)|^2$ and correlation function of a stationary field are a Fourier pair.
That is,
\begin{equation}
\langle f(\vr)f(\vr+\Delta\vr)\rangle
=\int\d^3\vk\,\e^{\i\vk\cdot\Delta\vr}|\tilde f(\vk)|^2,
\label{eqn:WK_theorem}
\end{equation}
where 
\begin{equation}
  \begin{split}
\tilde f(\vk)
&\equiv \frac{1}{(2\pi)^{3/2}}
\int\d^3\vr\, \e^{-\i\vk\cdot\vr}f(\vr)\\
  \end{split}
\end{equation}
is the Fourier transform of $f(\vr)$.

If we modulate a stationary field $f(\vr)$ by multiplying it by a slowly varying function $a(\vr)$, the resulting $g(\vr)\equiv a(\vr)f(\vr)$ is an example of a {\it semi-stationary} process \citep{Priestley_only.1965}.   Then, instead of the Wiener--Khinchin theorem, we have that
\begin{equation}
  \begin{split}
  \label{eq:priestley}
  &\left\langle g(\vr_1)g(\vr_2)\right\rangle
  \simeq a(\vr_1) a(\vr_2)
  \int\d^3\vk\, \e^{\i\vk\cdot(\vr_2-\vr_1)}|\tilde f(\vk)|^2.
  \end{split}
\end{equation}
This is a good approximation when the Fourier transform $\tilde a(\vk)$ of the modulating function has an absolute maximum at $\vk=0$. 

Finally, $f(\vr)$ is a {\it Gaussian random field} if, for any choice of $N$ points $\vr_1$,...,$\vr_N$, the joint pdf $\pr(\vf)$ is a multivariate normal.  That is,
\begin{equation} \pr(\vf)\,\d\vf=\frac1{(2\pi)^{3N/2}|\bm{\Sigma}|^{N/2}}\exp\left[-\frac12(\vf-\bar
  \vf)^T \bm{\Sigma}^{-1}(\vf-\bar\vf)\right]\,\d\vf,
\label{eq:GRFdefn}
\end{equation}
in which the mean values $\bar\vf\equiv(\bar f(\vr_1),...,\bar f(\vr_N))$ and the covariance matrix $\Sigma_{ij}\equiv\cov[f(\vr_i),f(\vr_j)]$ depend only the location of the points $\{\vr_n\}$.
Such a field is completely specified by its mean $\bar f(\vr)$ and covariance $\cov[f(\vr_i),f(\vr_j)]$ functions.
As the correlation function is related to the mean and covariance through
\begin{equation}
 \langle f(\vr)f(\vr+\Delta \vr)\rangle = \bar f^2+\mathcal{C}(|\Delta\vr|),
\end{equation}
it follows that a stationary Gaussian random field is completely determined by its power spectrum~$|\tilde f(\vk)|^2$.
If $f=\log A$ is a Gaussian random field, then $A(\vr)=\e^{f(\vr)}$ is {\it not} a Gaussian random field, as the joint probability density of $A(\vr_1),...,A(\vr_N)$ does not satisfy the condition~\eqref{eq:GRFdefn}.
Instead, any field $A(\vr)$ for which $\log A$ is a Gaussian random field is called a {\it lognormal} random field.

\subsubsection{Length scales}\label{sec:scales}

On the largest scales we do not expect the dust density to be a stationary process; clearly there are large regions where one expects the dust density to be relatively high (e.g., along spiral arms) and other regions where it will be low.
We can, however, split our problem into two regimes by spatial scale: on larger scales the mean function of dust density will vary slowly with position, describing features such as spiral arms; whilst on smaller scales the turbulent ISM is resolved with a Gaussian random field.
The question then becomes at which scale should the regimes be split and why?

We can address this by considering the power spectrum of dust density, which is generally thought to follow a power law across an inertial range between inner and outer limiting scales.
The outer scale is determined by the dominant means of energy injection into the medium, with reduced power on larger scales.
It is believed that the injection of energy into the ISM is dominated by supernovae, which produce features on spatial scales of the order of $100$~pc \citep{Elmegreen_Scalo.2004, MacLow_only.2004}.
Many other processes inject energy at larger or smaller scales, but are thought to add less energy than supernovae.
Although the outer scale has not been identified in observations of Galactic dust, \cite{Haverkorn_Brown.2008} and \cite{Chepurnov_Lazarian.2010} find that in interarm regions the outer scale of fluctuations in the warm ionised and neutral medium is on the order of $100$~pc.
Meanwhile, the power spectra of dust emission in M33 \citep{Combes_Boquien.2012} is broken at a similar scale.
\cite{Schlegel_Finkbeiner.1998} found no clear break in the power spectrum of Galactic dust emission, but we note that this is most likely a result of the projection of the 3D Galaxy onto a 2D image.

At small scales the onset of viscosity dominance reduces power and the ISM becomes essentially smooth.
The inner scale is believed to be rather small.  In the ionised interstellar medium it has been measured to be on the order of 100~km \citep{Spangler_Gwinn.1990}.
There exist no observations of interstellar dust which probe such fine scales, but it is clear that it does exhibit structure on scales significantly smaller than a parsec \citep[e.g.][]{DiFrancesco_Sadavoy.2010, Robitaille_Joncas.2014}.
Therefore, the inner scale of turbulence can be safely ignored in this context, as we operate on significantly coarser scales.

\subsubsection{Covariance function of $\rho$}\label{sec:cov_rho}

Our model for the dust density is that $\rho$ is a semi-stationary random field, with slowly varying mean $m_\rho(\vr)$ and standard deviation $S_{\rho}(\vr)$.
For example, this model can be obtained by expressing the logarithm of the density as
\begin{equation}
  \log \rho(\vr)=a_\rho(\vr)+b_\rho(\vr)z(\vr),
\end{equation}
where $z(\vr)$ is a stationary random field having zero mean and unit variance.  The relationship between $(a_\rho(\vr),b_\rho(\vr))$ and $(m_\rho(\vr),S_\rho(\vr))$ depends on the distribution of $z(r)$.
If we assume that $z(r)$ is Gaussian, then
\begin{align}
m_{\rho}(\vr) &=  e^{a_\rho(\vr)+b_\rho^2(\vr)/2}, \\
S_{\rho}^2(\vr) &=  ( e^{b_\rho^2(\vr)} -1 )e^{2a_\rho(\vr)+b_\rho^2(\vr)}. 
\end{align}
In Appendix~\ref{app:test} we show that this assumption produces a $\log A$ distribution that is very close to Gaussian.  We emphasise, however, that our method assumes only that $\log A$ is a Gaussian random field; the details of the $\rho$ field that produces this lognormal extinction distribution are not important, provided only that $\rho$ is some semi-stationary random field, described by some mean $m_\rho(\vr)$, variance~$S_\rho(\vr)$ and power spectrum.

Following the considerations outlined in Section~\ref{sec:spatial_scaling}, we want the density power spectrum to have the Kolomogorov-like form~\eqref{eq:kolmopower}.
But, as noted in Section~\ref{sec:scales}, this scale-free form cannot hold for all wavenumbers $|\vk|$.
So, we take the power spectrum of $\rho$ to be proportional to the function
\begin{align}
\Upsilon_{\gamma, \Omega}(\vk, \mathcal{L}_o)
\equiv
 R(\gamma, \Omega,\mathcal L_o) \frac{ (|\vk| \mathcal{L}_o )^{2\Omega} }{ \left[ 1+ (|\vk| \mathcal{L}_o )^2 \right]^{ \frac{\gamma}{2} + \Omega} },
\label{eq:rhopowerspec}
\end{align}
in which the normalization constant $R(\gamma,\Omega,\mathcal{L}_o)$ is given by
\begin{align}
\frac{1}{R(\gamma, \Omega,\mathcal{L}_o)} = 4\pi \int_0^{\infty} d|\vk| \frac{ |\vk|^2 (|\vk| \mathcal{L}_o )^{2\Omega} }{ \left[ 1+ (|\vk| \mathcal{L}_o )^2 \right]^{ \frac{\gamma}{2} + \Omega} },
\end{align}
so that taking $|\tilde\rho(\vk)|^2=\Upsilon_{\gamma,\Omega}(\vk,\mathcal{L}_0)$ results in a density field with unit variance.
We then scale the spectrum by $S^2_{\rho}(\vr)$ to account for variations in the standard deviation of the field and add a $\delta$-function at the origin, scaled by $m^2_\rho(\vr)$, to encode variations in the mean density.
This spectrum has a Kolmogorov-like $|\vk|^{-\gamma}$ shape within the characteristic scalelength $\mathcal{L}_o$.
On larger scales the spectrum tapers smoothly to zero, the shape of the taper being set by the parameter~$\Omega$.
By tapering the spectrum, structures on scales greater than $\mathcal{L}_o$ are described only by variations in $m_{\rho}(\vr)$ and $S_\rho(\vr)$ and are unaffected by the random field~$z(\vr)$.
This gives rise to the split in scales discussed in section~\ref{sec:scales}.
We do not adjust~\eqref{eq:rhopowerspec} to account for the inner scale of turbulence, as this is significantly smaller than any of the scales we can probe in the ISM.

\begin{figure*}
\includegraphics{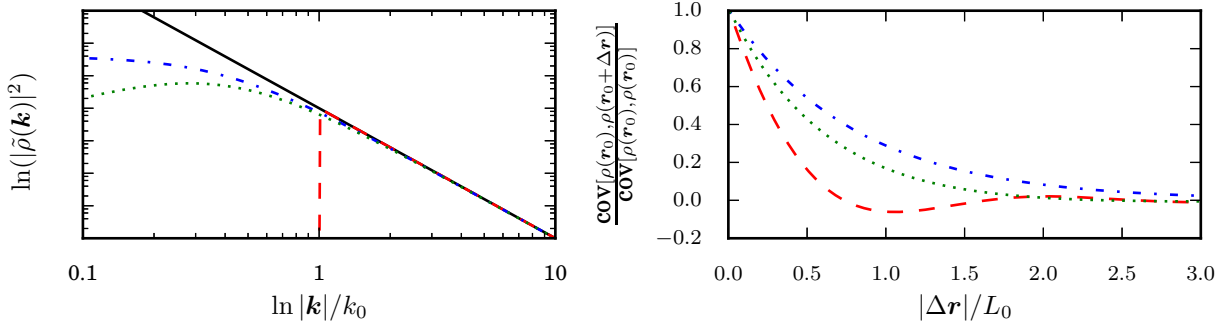}
\caption{A comparison of the power spectra (left) and normalised covariance functions (right) with four different forms of truncation at large scales. The black solid line shows the untruncated form, the red dashed line shows sudden truncation at $k_0$, whilst the $\Upsilon_{11/3, 0}$ and $\Upsilon_{11/3, 1}$ forms are shown with blue dot-dashed and green dashed lines respectively. 
\label{fig:PS-autcorr}}
\end{figure*}

We note that, as indicated by equation~\eqref{eqn:WK_theorem}, setting the power spectrum of $\rho(\vr)$ is equivalent to setting its covariance function, since the two are a Fourier pair.
However, setting the power spectrum is more convenient, because it offers a straightforward way of employing a physically motivated Kolmogorov-like spectrum.
Fig.~\ref{fig:PS-autcorr} compares power spectra of different forms and the covariance function pairs of the power spectra considered.

\subsubsection{Covariance function of $\log A$}

We model $\log A$ as a Gaussian random field whose mean and covariance functions depend on the density function above.
The mean is straightforward to calculate from the function $m_\rho(\vr)$, given the covariance function of $\log A$.
The relationship between the covariance of $\log A$ and the density is less simple and so we relegate the details to Appendix~\ref{app:covA_gen}.
Equation~\eqref{eq:covlogA} gives the covariance function of $\log A$ which depends on the power spectrum~\eqref{eq:rhopowerspec} we choose for the dust density as well as on $m_\rho(\vr)$ and $S_\rho(\vr)$, but does not depend on the form of the distribution of $\log \rho$.

\section{Constructing maps}\label{sec:mapmaking}

The problem we address in this paper is that of inferring the three-dimensional distribution of extinction from some data.
Typically these data will come in the form of a catalogue or catalogues of observations of some set of stars, whether photometric, spectroscopic, astrometric, or some combination of the three.
For example, one could employ 2MASS photometry on its own, or use 2MASS combined with IPHAS photometry, or 2MASS plus a spectroscopic survey, such as SEGUE, RAVE or APOGEE.
Subsequently, one might want to refine the maps by including parallaxes from Gaia as they become available.

We split the problem into three tasks.
First, given some catalogue of observations of stars, we would like to estimate the extinction towards each of the stars in the catalogue: as discussed in \cite{Sale_only.2012}, a properly constructed, coherent extinction map also helps us determine other properties of these stars -- such as their distance or masses -- more precisely than would be possible if the stars were studied separately.
Second, from this catalogue we would like to infer the extinction to arbitrary positions in space.
For example, we could be interested in the extinction to stars or other objects that were not included in the catalogue used to construct the map.
Finally, we seek to characterise the distribution of extinction and, by extension, dust.
We are interested in how they are distributed on large scales as well as the detailed statistical behaviour of dust which informs us about the turbulent processes that shape the ISM.

In the following we introduce a barrage of labels to denote various parameters and modelling assumptions.  These are summarised in Table~\ref{tab:param_sum}.

\begin{table*}
\begin{tabular}{c|l}
$(\tilde l_n, \tilde b_n)$ & the $(l,b)$ coordinates observed for the $n^{\rm th}$ star in the catalogue.\\
$\tilde{y}_n$ & all other directly observed quantities (e.g., broad-band fluxes, trigonometric parallax)
for the $n^{\rm th}$ star.\\
\noalign{\smallskip}
$s_n$ & the distance to the $n^{\rm th}$ star in the catalogue. \\
$A_n$ & the extinction to the $n^{\rm th}$ star in the catalogue. \\
$x_n$ & all other intrinsic properties of the $n^{\rm th}$ star (mass, metallicity, age, etc).\\
\noalign{\smallskip}
$\macromodel$ & parameters that set the large-scale distribution of extinction, $m_\rho(\vr)$, $S_{\rho}(\vr)$ \\
$\macroasumps$ & any assumptions made about $\macromodel$ (e.g., an assumed functional form for $m_\rho(\vr),S_{\rho}(\vr)$) \\
$\micromodel$ & the model for the small scale distribution of $\log \rho$, including $(\gamma,{\cal L}_0,\Omega)$\\
$\galaxymodel$ & the background Galaxy model, including prior on position, metallicity, age, etc. \\
$\transfermodel$ & denotes model for mapping $(s_n,A_n)$ to predicted observables, $\bar y_n$.\\
\end{tabular}
\caption{A list of the parameters and assumptions used in our model.
\label{tab:param_sum} }
\end{table*}

\subsection{Probability of a map}

We use $\macromodel$ to stand for the set of parameters of the functions $m_{\rho}(\vr)$ and $S_{\rho}(\vr)$ that describe the large-scale distribution of extinction.
These parameters could be as simple as a scale height and scale length describing an exponential disc.
A more sophisticated model might use three-dimensional splines, in which case $\macromodel$ would contain the positions of the spline points, the values of $m_\rho(\vr)$ and $S_\rho(\vr)$ at each point, and any prior assumptions about the covariances among the values.
We use $\micromodel$ to denote our model for the small-scale properties of the ISM (such as $\gamma$, $\mathcal{L}_0$ and $\Omega$) and $\transfermodel$ for our model for {\it how} stars, subject to some extinction, would appear in observations.
For example, a simple model for the latter may include isochrones, synthetic stellar atmospheres and a model for the wavelength dependence of extinction.

We are given a catalogue of $N$ stars.
Let $(\tilde l_n,\tilde b_n)$ be the observed Galactic coordinates and let $\tilde{y}_n$ be the full set of all other observations (observed parallax, apparent magnitudes, etc) of the $n^{\rm th}$ star in the catalogue.
We use the tilde to indicate that these are observations and not the true values.
Each star has some unknown distance, $s_n$, and extinction, $A_n$, along with a set of other intrinsic parameters (mass, age, metallicity and so on) that we denote by $x_n$.
We use vectors $(\setoflhat,\setofbhat)=(\{\tilde l_n\},\{\tilde b_n\})$ and $\setofyhat=\{\tilde{y}_n\}$ to denote the full set of observations in the catalogue.
Similarly, $\setofs=\{s_n\}$ and $\setofA=\{A_n\}$ refer to the full set of distances and extinctions to each star in the catalogue, while $(\setofl,\setofb)=(\{l_n\},\{b_n\})$ denote the true Galactic coordinates of the stars in the catalogue.

We assume a prior model, $\galaxymodel$, for the joint probability distribution of the parameters $(l_n,b_n,s_n,x_n)$ describing each star's three-dimensional position, age, metallicity, and so on.
We treat all but the position of the star and its extinction as nuisance parameters and use the method of \citet{Sale_only.2012}, or equivalently \citet{Hanson_Bailer-Jones.2014} or \citet{Green_Schlafly.2014}, to marginalise  $x_n$ to obtain the likelihood $\pr( \tilde l_n,\tilde b_n,\tilde{y}_n|l_n,b_n,s_n,A_n,\galaxymodel,\transfermodel)$ of $(l_n,b_n,s_n,A_n)$ for each star given the assumed physical model~$\transfermodel$ for the dust extinction and the background galaxy model~$\galaxymodel$.  
Examples of these marginalised likelihoods can be seen in Figure~6 of \citet{Green_Schlafly.2014}.

Using this likelihood, the posterior distribution
\begin{equation}
  \begin{split}
&\pr(\macromodel, \setofl,\setofb, \setofs,\setofA| \setoflhat, \setofbhat,
\setofyhat, \micromodel, \transfermodel,
\galaxymodel, \macroasumps)\\
&\quad \propto \pr(\setofA|\setofl,\setofb,\setofs,\macromodel,\micromodel)\pr(\macromodel|\macroasumps)\\
&\qquad\times\prod_n\pr(\tilde l_n,\tilde b_n, \tilde{y}_n|l_n,b_n,s_n, A_n, \transfermodel,\galaxymodel)
\pr( l_n, b_n, s_n,|\galaxymodel).
  \end{split}
\label{eqn:model_theta}
\end{equation}
Here $\pr(\setofA|\setofl,\setofb,\setofs,\macromodel,\micromodel)$ is a multivariate
Gaussian on $\log \setofA$, the covariance matrix of which is set
using the covariance function $\cov[\log A(l_1,b_1,s_1),\log
A(l_2,b_2,s_2)]$ given by equation~\eqref{eq:covlogA}, which in turn depends on an assumed dust power spectrum as in equation~\eqref{eq:rhopowerspec}.
The mean vector for $\log \setofA$ is found by first integrating the mean density along the line of sight to each star to find the expected extinction to each star.
By combining these and the covariance matrix it is then possible to find the expectation of log-extinction for all the stars.

We assume that our observations of the on-sky position of each star are independent of the other observations of that star:
\begin{equation}
  \begin{split}
&\pr(\tilde l_n,\tilde b_n,\tilde{y}_n|l_n,b_n,s_n,A_n, \transfermodel,\galaxymodel)\\
&\qquad = \pr(\tilde l_n,\tilde b_n|l_n,b_n,\transfermodel)
\pr(\tilde{y}_n|l_n,b_n,s_n, A_n, \transfermodel,\galaxymodel).
  \end{split}
\end{equation}
Typically the uncertainties in $(\tilde l_n,\tilde b_n)$ are negligible and we may take 
\begin{align}
\pr(\tilde l_n,\tilde b_n|l_n,b_n,\transfermodel)
=\delta(\tilde l_n-l_n)\delta(\tilde b_n-b_n).
\end{align}
Marginalising $(\setofl,\setofb)$, equation~\eqref{eqn:model_theta} simplifies to
\begin{equation}
  \begin{split}
&\pr(\macromodel, \setofs, \setofA| \setoflhat,\setofbhat, \setofyhat, \micromodel, \transfermodel,
\galaxymodel, \macroasumps) \\
&\quad = \pr(\setofA|\setoflhat,\setofbhat,\setofs,\macromodel,\micromodel)
\pr(\macromodel|\macroasumps)\\
&\qquad\times
\prod_n
\pr(\tilde{y}_n|\tilde l_n, \tilde b_n,s_n, A_n, \transfermodel,\galaxymodel)
\pr(s_n| \tilde l_n,\tilde b_n,\galaxymodel).
  \end{split}
\label{eq:model}
\end{equation}
In $\pr(\macromodel|\macroasumps)$ we include any assumptions we wish to make about the parameters that define the large-scale structure of extinction.
For example, a simple model for the dust density might include a characteristic scale height among the parameters~$\macromodel$.  Then the prior assumed for this scale height would be incorporated into $\pr(\macromodel|\macroasumps)$. 
Finally, $\pr(s_n|\tilde l_n,\tilde b_n,\galaxymodel)$ is a prior on the distance to the stars based on the background model~$\galaxymodel$.

\subsection{Extinction and distance to a single star in the catalogue}\label{sec:marg_star}

If we are interested in a particular star, we can obtain the marginal posterior for only this star by marginalising over the parameters of the other stars and $\macromodel$:
\begin{align}
\begin{split}
&\pr(s_1,A_1| \setoflhat,\setofbhat, \setofyhat,\micromodel, \transfermodel, \galaxymodel, \macroasumps) \\
&\quad = \int \pr(\macromodel, \setofs, \setofA| \setoflhat,\setofbhat,\setofyhat,\micromodel, \transfermodel, \galaxymodel, \macroasumps) \, \d\setofs_{2\cdots N} \, \d\setofA_{2\cdots N}\,\d\macromodel \\
\end{split} \\
\begin{split}
&\quad = \pr(\tilde{y}_1|\tilde l_1,\tilde b_1,s_1,A_1,\transfermodel,\galaxymodel)
\pr(s_1 |\tilde l_1,\tilde b_1,\galaxymodel)\\
&\qquad\times  \int \d\macromodel\int\d\setofs_{2\cdots N}
\int\d\setofA_{2\cdots N}\Bigg[
\pr(\setofA|\setoflhat,\setofbhat,\setofs,\macromodel,\micromodel)
\pr(\macromodel|\macroasumps)\\
&\qquad\qquad
\times\prod_{n=2}^{N}
\pr(\tilde{y}_n|\tilde l_n,\tilde b_n,s_n, A_n, \transfermodel,\galaxymodel)\pr(s_n|\tilde l_n,\tilde b_n,\galaxymodel)
\Bigg].
\label{eqn:marg_post}
\end{split}
\end{align}
Notice how all the other stars influence the posterior of the star we are interested in. If we took this star to be independent of all the others the posterior would reduce to the product of the first two terms on the right hand side of equation~\eqref{eqn:marg_post}.  By using the extra information given by the observations of the other stars, however, it is possible to obtain a more precise estimate of the distance and extinction to the star than if it had been considered in isolation.\footnote{Of course, if our model for the covariance of $\log A$ is wrong, our inferred $A_n$ and~$s_n$ -- however precise --  will be biased.}
It might appear that such an integration above would be expensive to perform, but it is naturally obtained from the MCMC algorithms we use to sample the posterior~\eqref{eq:model}; a sample from the marginalised posterior on $s_1$ and $A_1$ is found by only considering these variables in the MCMC chain.

\subsection{Inferring extinction to arbitrary position(s)}\label{sec:posterior_pred}

Another question one might ask is, what is the extinction to an arbitrary position or positions?
For example, one might want to know the extinction to a non-stellar object or to a star that did not appear in the original catalogue.
We can obtain the extinction $A^{\star}$ to any position $(l^\star,b^\star,s^\star)$ from the posterior predictive distribution,
\begin{equation}
  \begin{split}
&\pr(A^\star|l^\star,b^\star,s^\star, \setoflhat,\setofbhat, \setofyhat,  \micromodel, \transfermodel, \galaxymodel, \macroasumps)\\
& = \int     \,\d\setofA \,\d\setofs \, \d\macromodel\\
&\qquad\qquad
\pr(A^\star|l^\star,b^\star,s^\star,\setoflhat,\setofbhat, \setofs,\setofA, \macromodel,\micromodel)
\pr(\macromodel, \setofs, \setofA|\setoflhat,\setofbhat, \setofyhat, \micromodel, \transfermodel, \galaxymodel, \macroasumps).
  \end{split}
\label{eq:singleAstar}
\end{equation}
This marginalisation is most straightforwardly carried out by using an MCMC algorithm to explore the posterior distribution $\pr(\macromodel, \setofs, \setofA|\setoflhat,\setofbhat, \setofyhat, \micromodel, \transfermodel, \galaxymodel, \macroasumps)$ and summing up the samples to obtain a realisation of $\pr(A^\star|l^\star,b^\star,s^\star,\setoflhat,\setofbhat, \setofs, \setofA, \macromodel,\micromodel)$.
The posterior is obtained directly from our assumption that~$\log A$ is Gaussian random field.  This lognormal assumption means that finding the joint probability distribution of the extinction to multiple points in space is an obvious, straightforward generalisation of~\eqref{eq:singleAstar}

\subsection{General structure of dust}

One can also obtain the marginal posterior of the large-scale structure of dust, $\pr(\macromodel|\setoflhat,\setofbhat,\setofyhat,\micromodel,\transfermodel,\galaxymodel)$, by
marginalising~$\setofs$ and $\setofA$ from equation~\eqref{eq:model}.  As in section~\ref{sec:marg_star}, this can be easily achieved if an MCMC algorithm has been employed.

We postpone to a future paper the question of how to infer the parameters $(\gamma,\mathcal{L},\Omega)$ that govern the model~$\micromodel$ for the small-scale physics.  For now we simply assume that $\gamma=11/3$, $\mathcal{L}_0=100$~pc and $\Omega=1$.

\section{Examples}\label{sec:2d}
\def\true{,\rm true}

We now consider some examples to illustrate how to use the formalism above to construct extinction maps within a narrow, two-dimensional wedge of a model galaxy.  
We restrict ourselves to two dimensions only because it is then straightforward to present the results on paper; there is nothing fundamentally different in our scheme between this simple two-dimensional wedge and the full three-dimensional case.
As we work in two dimensions we can describe a star's position with its distance $s_n$ and Galactic longitude $l_n$ only, enabling us to drop the Galactic latitude $b_n$.

To focus attention on the dust model, the model galaxy~$\galaxymodel$ we consider here is deliberately crude: there is a single population  of stars, the spatial density of which is uniform within the wedge, so that 
\begin{align}
\pr(s_n|l_n, \galaxymodel) \propto s_n.
\end{align}
 The dust density model~$\macromodel$ is slightly more realistic.  The function that sets the mean density of the dust distribution 
falls off exponentially with distance from the observer:
\begin{equation}
\bar\rho(l,s) \equiv
m_\rho(l,s) =\rho_0 \exp \left(- \frac{s}{s_l} \right).
\label{eq:testdustdensity}
\end{equation}
This is intended to mimic the situation in which one is modelling observations towards the Galactic anticentre $(l,b)=(180^\circ,0^\circ)$.
We take the ratio between the standard deviation and mean of dust density to be constant $S_{\rho}(l,s)/m_\rho(l,s)=\sqrt{\e-1}$, where $\e$ is the base of natural logarithms.
Therefore our model~$\macromodel$ describing the large-scale dust density has only two free parameters, $\rho_0$ and $s_l$.
In the following we adopt a uniform prior on $\log \rho_0$ and $\log s_l$.

We avoid specifying an explicit model~$\transfermodel$ for the
extinguishing effects of dust on the observables $\tilde y_n$ by taking
$\tilde y_n=(\log\tilde s_n,\log\tilde A_n)$ and assigning
observational uncertainties to the ``observed'' (log)
distance, $\log\tilde s_n$, and ``observed'' (log) extinction, $\log\tilde A_n$, to
each star.

To construct our simulated galaxy catalogue we first use the procedure 
described in appendix~\ref{sec:simulations} to
generate a three-dimensional realisation of the dust density
with power spectrum proportional to~\eqref{eq:rhopowerspec}, which are modulated by the mean density $m_\rho(s)$ given by equation~\eqref{eq:testdustdensity} having scale length 
$s_l=2\,\rm kpc$ and asymptotic extinction
\begin{equation}
  \label{eq:A0}
 A_\infty\equiv \kappa\rho_0 s_l=5.0.
\end{equation}
Then we sprinkle 200 stars on a wedge of length 10~kpc and opening angle $30\arcmin$ within this volume.  
We read off the true distances and extinctions $y_{n, {\rm true}}=(s_{n\true},A_{n\true})$ to each star, then scatter each by an amount consistent with the assumed observational uncertainty to obtain the ``observed'' values $\tilde y_n$.

\begin{figure}
\includegraphics{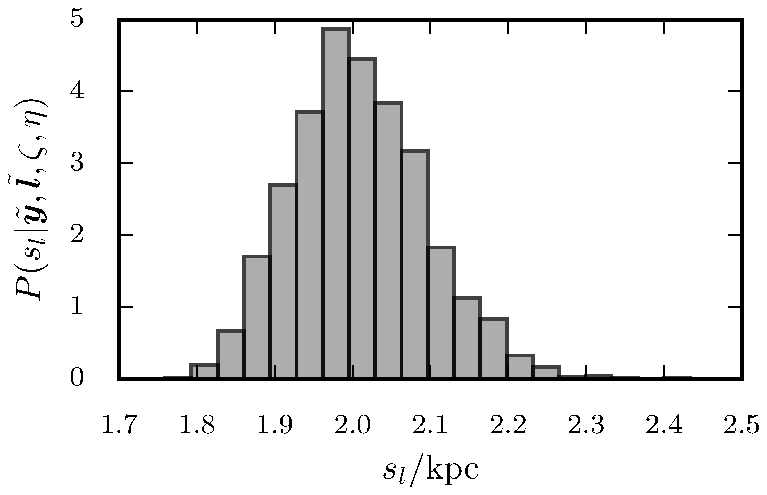}
\caption{The marginal posterior distribution of dust scale length $s_l$ found from perfect two-dimensional data. The true value of $s_l$ used to simulate the data is $s_l=2$~kpc. \label{fig:d0_post_perf_2d}}
\end{figure}

\begin{figure}
\includegraphics{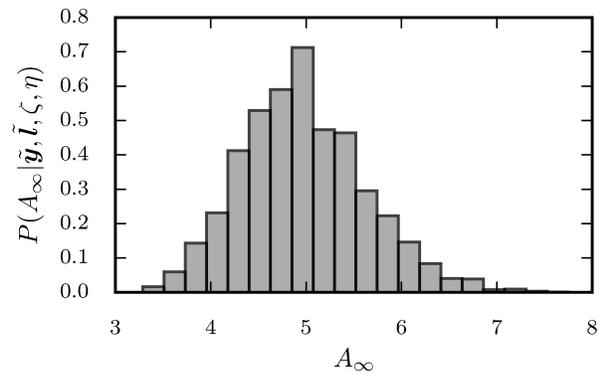}
\caption{The marginal posterior distribution of asymptotic extinction $A_\infty$ found from perfect two-dimensional data. The true value of $A_\infty$ used to simulate the data is $A_\infty=5.$. \label{fig:a0_post_perf_2d}}
\end{figure}

\begin{figure*}
\includegraphics{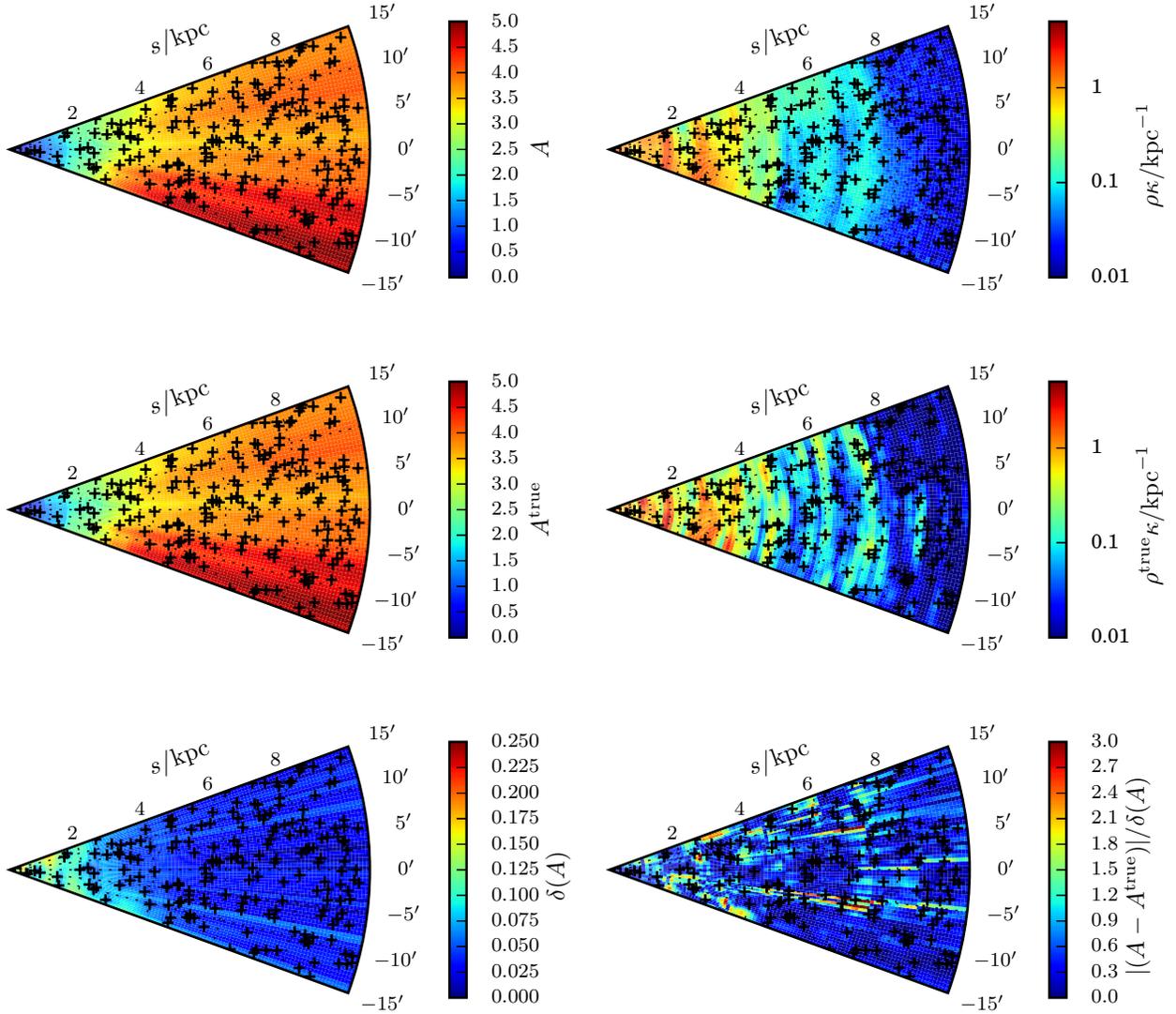}
\caption{A series of maps demonstrating the results our method achieves from `perfect' simulated data. We plot the posterior expectation of extinction (top left) and density (top right), the true extinction (middle left) and density (middle right), the width of a 68\% credible interval on extinction as a measure of uncertainty (bottom left) and the residual between the posterior expectation of extinction and true extinction normalised by the uncertainty. On all the plots crosses indicate the positions of the stars `observed'.
 \label{fig:Ad_perf_2D}}
\end{figure*}

\begin{figure*}
\includegraphics{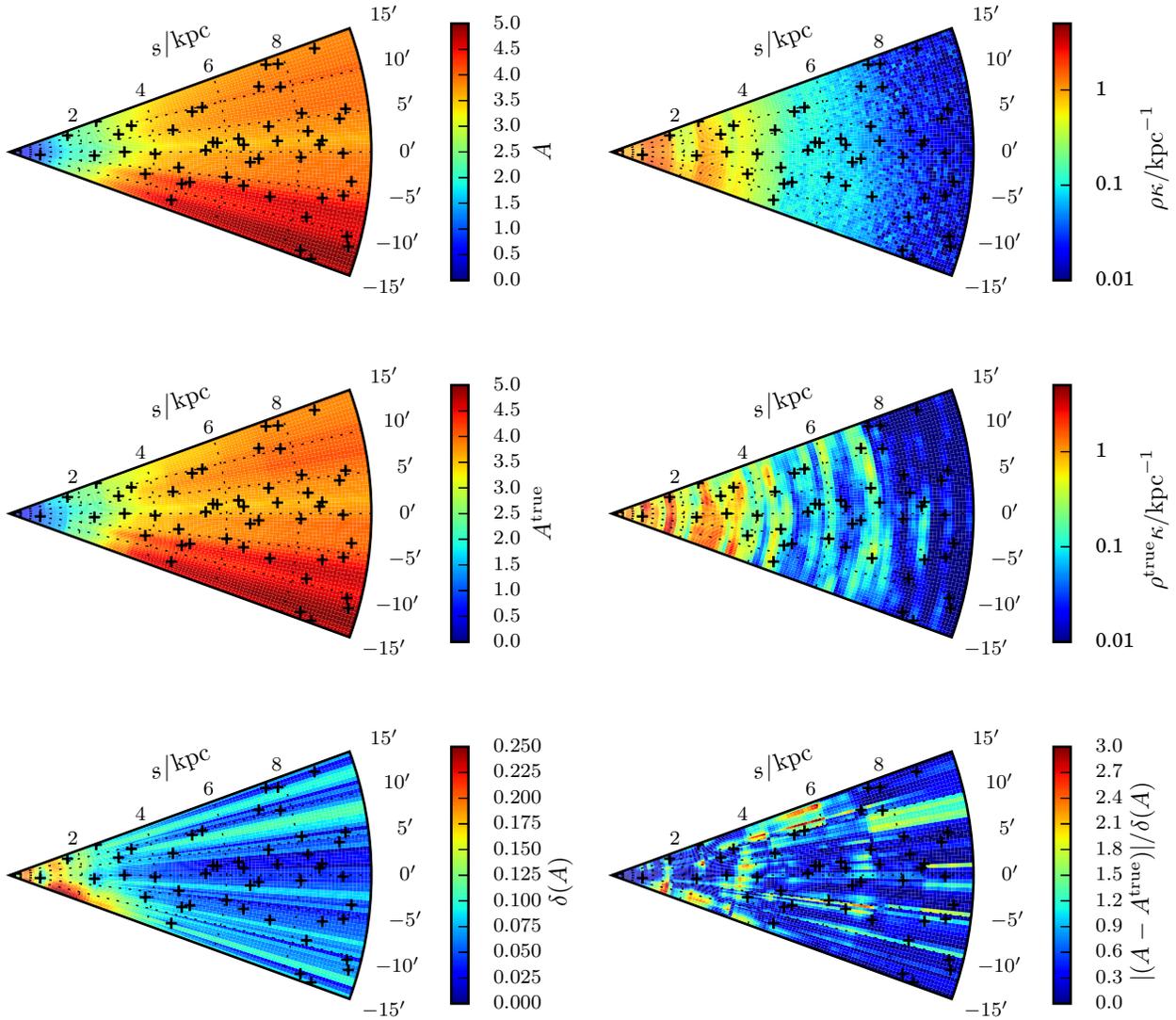}
\caption{As with Fig~\ref{fig:Ad_perf_2D}, but only employing 50 of the original 200 stars.
 \label{fig:Ad_perf_2D_reduced}}
\end{figure*}

\subsection{Perfect Data}\label{sec:2d_perf}

We first consider the case where we possess perfect data, meaning that we know the parameters of the stars with absolute precision.
Then 
\begin{equation}
\begin{split}
&\pr(\setofyhat |\setoflhat, \setofs, \setofA,
  \transfermodel) \pr(\setofs | \setoflhat, \galaxymodel)\\
&\quad =
\prod_n\delta(\log s_n-\log s_{n\true})\,
\delta(\log A_n-\log A_{n\true})
.\label{eq:deltafnlik}
\end{split}
\end{equation}
The only uncertain parameters are those describing the mean
extinction, $\macromodel=(a_0,s_l)$, and, 
instead of considering the full posterior described by equation~\eqref{eq:model}, we need only consider
\begin{align}
\pr(\macromodel|\setofyhat, \setoflhat,\micromodel, \macroasumps)
\propto
\pr(\setofA_{\rm{true}} | \setoflhat, \setofs_{\rm{true}},
  \macromodel, \micromodel)
\pr(\macromodel | \macroasumps), \label{eqn:perfect_posterior}
\end{align}
where $\pr(\setofA_{\rm{true}} | \setoflhat, \setofs_{\rm{true}}, \macromodel, \micromodel)$ is a Gaussian in $\log(\setofA_{\rm{true}})$, with a covariance matrix set using the covariance function described in appendix~\ref{app:covA_gen}.
This posterior can be estimated using a simple Markov chain Monte Carlo (MCMC) algorithm to sample the members of $\macromodel$.
We run an MCMC algorithm with a Metropolis-Hastings updater for $\macromodel$ for $20000$ iterations, discarding the first $5000$ iterations in the chain as burn-in. 
Both the overall length of the chain and the length of burn in are rather conservative: the autocorrelation length of the chain is less than 40 iterations and the chain typically converges within the first thousand iterations.

The marginal posteriors on $s_l$ and $A_\infty$ for an example realization are shown in Figures~\ref{fig:d0_post_perf_2d} and~\ref{fig:a0_post_perf_2d}.
Subsequently we continue by calculating the posterior predictive distribution $\pr({A}^{\star} | l^\star,s^{\star}, \setoflhat,\setofyhat, \transfermodel, \galaxymodel, \micromodel, \macroasumps )$.
This requires marginalising over $\macromodel$, which is achieved simply by iterating over the different members of the MCMC chain for $\macromodel$.
The extinction map produced in this way is shown in Figure~\ref{fig:Ad_perf_2D}.

The resolution of the map is fundamentally limited by the distribution of stars.
To demonstrate this we remove 150 stars from the catalogue used in this section and reanalyse.
Figure~\ref{fig:Ad_perf_2D_reduced} shows the extinction map obtained.
There is a clear loss of detail, particularly in the density map and a small increase in uncertainty.
None the less the larger extinction features are still found.

\subsection{Imperfect data}\label{sec:2d_imp}

Next we consider the effects of observational uncertainties by adding 
random errors to the true $(\vs,\vA)$ as follows.  For each star we
draw $\Delta y_n$ 
from a Gaussian having zero mean and covariance
\begin{align}
\bm{\Sigma}_n=
\begin{pmatrix}
\Sigma_{\log s,n} & \tau_n\sqrt{\Sigma_{\log s,n}\Sigma_{\log A,n}}\\
\tau_n\sqrt{\Sigma_{\log s,n}\Sigma_{\log A,n}} &\Sigma_{\log A,n}
\end{pmatrix}.
\end{align}
The variable $\tau_n$ is a random number drawn from the uniform distribution $[-1,1]$ and is included to model correlations between the observations of $s_n$ and $A_n$ for a single star.  
Note that we do not include any star--star correlations: $\Delta y_n$ is independent of $\Delta y_m$ for $m\ne n$.  
We model $\Sigma_{\log A,n}$ in a such a way that it grows as the extinction and/or distance to the star increases.
This is done by first giving each star a `true' apparent magnitude in the $V$ band,
\begin{align}
V_n = 2 + 5 \log_{10} (s_{n,{\rm true}}/10\textrm{pc}) + A_{n,{\rm true}} .
\end{align}
Uncertainties on extinction are then assumed to grow exponentially with apparent magnitude,
with variance
\begin{align}
\Sigma_{A,n} = e^{1.4 + 1.5 (V_i-23)} + 0.0025 ,
\end{align}
from which it is straightforward to find $\Sigma_{\log A,n}$.
We further set $\Sigma_{\log s,n} = 4 \Sigma_{\log A,n}$.
We set the uncertainties in this manner to imitate the growth of photometric uncertainties.
The specific scaling is applied to approximately match median uncertainties found from IPHAS data in \cite{Sale_Drew.2014}.
With this scaling, the rms scatter in $(\log s,\log A)$ varies from $\sim (0.06, 0.03)$ for a star at 2~kpc to $\sim (0.44,0.22)$ for one at 10 kpc.
We add this $\Delta y_n$ to $y_{n\true}$ to obtain our ``observations'',
\begin{align}
\tilde{y}_{n} = \begin{pmatrix}\log \tilde s_n \\ \log \tilde A_n \end{pmatrix} ,
\end{align}
where $\tilde s_n$ is the ``observed'' distance and $\tilde A_n$ the ``observed'' extinction to a star.

\begin{figure*}
\includegraphics{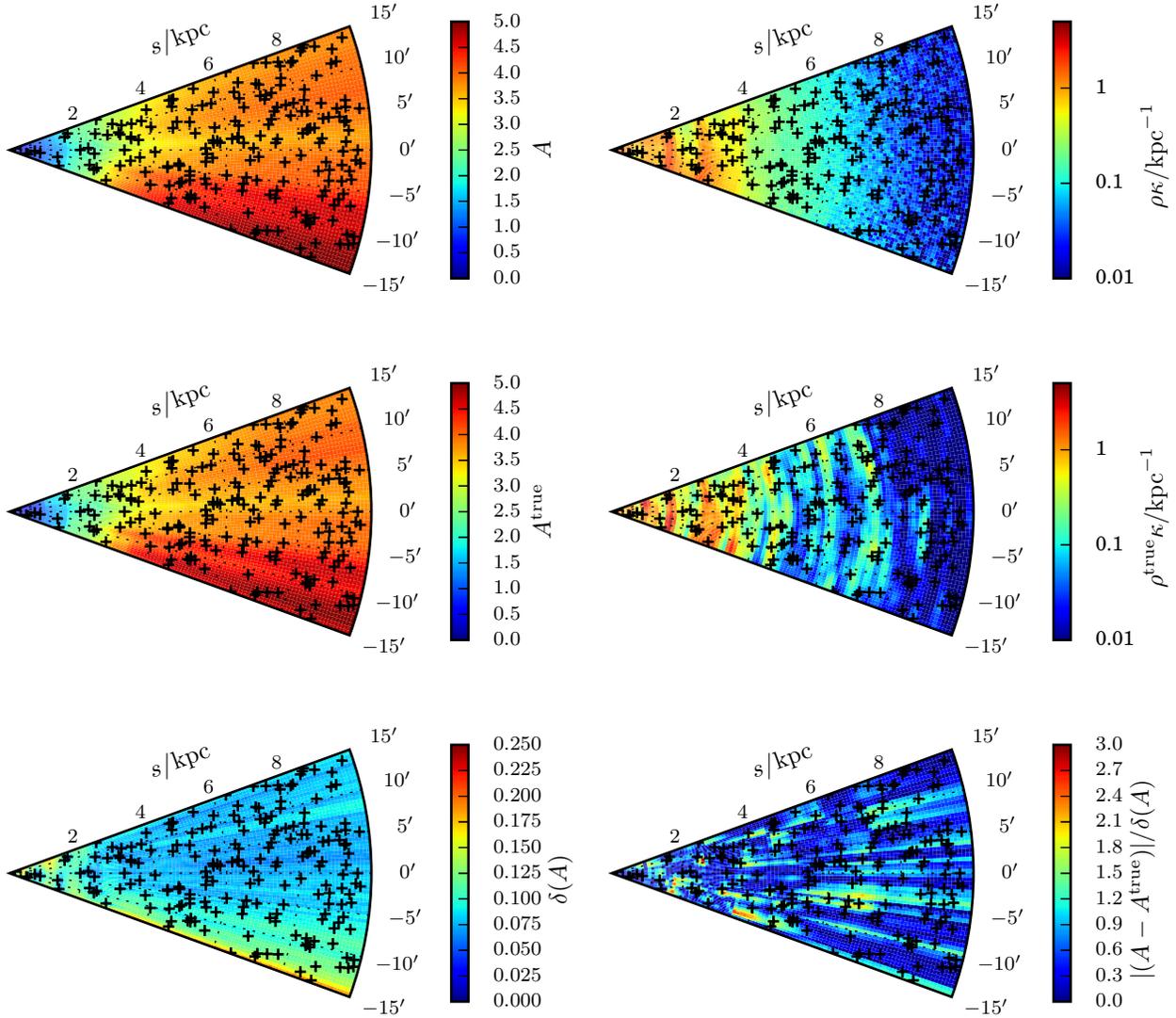}
\caption{As with Fig~\ref{fig:Ad_perf_2D}, but using data with simulated errors and uncertainties.
 \label{fig:Ad_imperf_2D}}
\end{figure*}
To model these observational uncertainties in our map-making
machinery we replace the the Dirac deltas in~\eqref{eq:deltafnlik} by the Gaussians
\begin{equation}
\begin{split}
&\pr(\setofyhat | \setoflhat, \setofs, \setofA, \transfermodel, \galaxymodel)  \pr(\setofs | \setoflhat, \galaxymodel)
\\
&\quad \propto \prod_n\frac1{2\pi|\bm{\Sigma}_n|^{1/2}}
\exp\left[-\frac12
  \left(\tilde y_n-\begin{pmatrix}\log s_n \\ \log A_n \end{pmatrix}\right)^T
  \bm{\Sigma}_n^{-1}
  \left(\tilde y_n-\begin{pmatrix}\log s_n \\ \log A_n \end{pmatrix}\right)
\right].
\label{eq:gaussianlik}
\end{split}
\end{equation}
With this it is straightforward to marginalise out $\setofA$ from the posterior equation~\eqref{eq:model}: for given $\setofs$, the integral of $\pr(\setofyhat |\setoflhat, \setofs, \setofA, \transfermodel, \galaxymodel) \pr(\setofA|\setoflhat,\setofs,\macromodel,\micromodel)$ with respect to $\log \setofA$ is simply a convolution of Gaussians.
Then we have that
\begin{equation}
  \begin{split}
    &
    \pr(\macromodel, \setofs | \setoflhat, \setofyhat,
    \transfermodel, \galaxymodel, \micromodel, \macroasumps)\\
&\quad\propto 
\pr(\tilde{\setofA} | \tilde{\setofs}, \setoflhat, \setofs,  \macromodel, \transfermodel, \micromodel)  \pr(\tilde{\setofs} | \setofs, \transfermodel)  \pr(\setofs | \setoflhat, \galaxymodel)  \pr(\macromodel | \macroasumps), \label{eqn:posterior_marg}
  \end{split}
\end{equation}
in which the first two factors are Gaussians in $\log\tilde\vA$ and $\log\tilde\vs$ respectively.
The covariance matrix for $\pr(\tilde{\setofA} | \tilde{\setofs}, \setoflhat, \setofs,  \macromodel, \transfermodel, \micromodel)$ is equal to that of $\pr(\setofA|\setoflhat,\setofs,\macromodel,\micromodel)$ plus the $\Sigma_{\log A,n}$ along the diagonal, whilst the mean vector is unchanged.

An obvious MCMC scheme for this posterior is to alternate updates of $\macromodel$ with updates to $\setofs$: this is an example of a Metropolis-within-Gibbs MCMC scheme \citep{Tierney_only.1994}.
We continue to employ a simple Metropolis-Hastings updater for $\macromodel$, whilst all members of $\setofs$ are updated simultaneously with another Metropolis-Hastings updater, as updating each member in sequence would require recalculating $\pr(\tilde{\setofA} | \tilde{\setofs}, \setoflhat, \setofs,  \macromodel, \transfermodel, \micromodel)$ for each member, which is computationally prohibitive.

The justification for marginalising over $\setofA$ follows from practical experience with this MCMC algorithm: the autocorrelation time in the marginalised case is $\sim 10$ times less than if $\setofA$ were also sampled.
If we desire, once we have used MCMC to estimate the posterior in equation~\eqref{eqn:posterior_marg}, we can then extend this to find the posterior on $\setofA$ using a similar process to that described in section~\ref{sec:posterior_pred}.

Figure~\ref{fig:Ad_imperf_2D} shows an example of the extinction and dust density maps reconstructed from our simulated catalogues.
The uncertainty in the extinction and density maps has increased relative to the case of perfect data, with some corresponding loss of detail, particularly in the density map.
However, the extinction map still largely reproduces the input.

\begin{figure}
\includegraphics{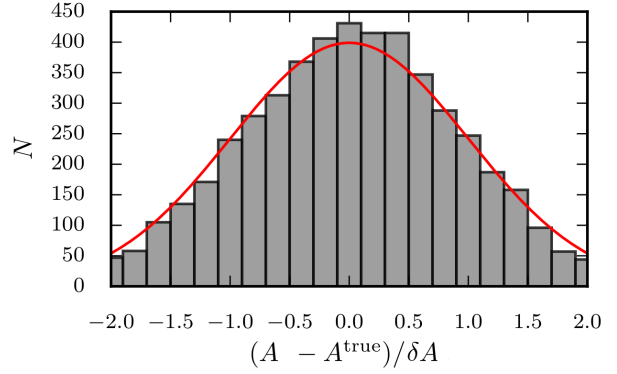}
\caption{A histogram of the posterior expectation of extinction to 5~kpc at an angle of $0\arcmin$ minus the `true' extinction to this position and divided by the standard deviation of the posterior. The measured mean of the data is 0.004 and the standard deviation 1.008.
Over plotted with a red line is a unit Gaussian.
 \label{fig:5000_comp}}
\end{figure}

To get a more rigorous handle on the reliability of our method we have simulated 5000 independent extinction fields and star catalogues, adding random observational errors to each.
We then analysed each as before.
To show an example of how we have used these to appraise the quality of our results we consider extinction along the line of sight $l=0~\arcmin$ to a distance of 5~kpc.
For each realization we have recorded the difference between the posterior expectation value of extinction to this position and the `true' value.
For comparison, we have also recorded the standard deviation of the posterior distribution, which serves to quantify the uncertainty in the posterior estimate.
Fig.~\ref{fig:5000_comp} shows the distribution of the ratio of these two quantities -- that is, the difference of the estimate from the true value divided by the uncertainty in the estimate.
The mean of the distribution shown is 0.004, confirming that the posterior expectation of extinction is a good estimator of the true extinction.
Additionally, the measured standard deviation of the plotted distribution is 1.008, suggesting that the standard deviation of the posterior is a reliable estimate of the uncertainty.

\begin{figure}
\includegraphics{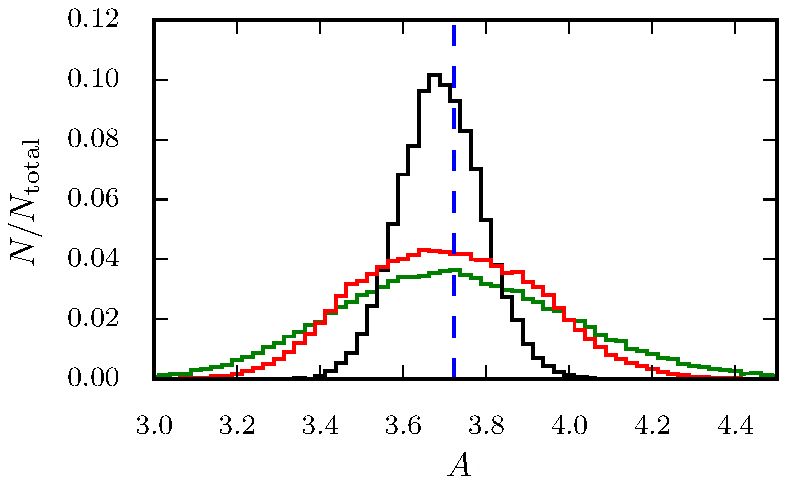}
\caption{Estimated posterior probability distributions of extinction for a single star located at 5169~pc.  The `true' extinction to the star is $A_0=3.72$, as indicated by the vertical dashed blue line. 
In black is a histogram of samples from $\pr(A_i | \setoflhat, \setofyhat, \micromodel, \galaxymodel, \transfermodel, \macroasumps)$, i.e. the posterior obtained from the method we describe once the distance of this star, the distance and extinction of all other stars and the large scale structure of dust have been marginalised over.
The red histogram represents a similar histogram obtained using the method of \protect\cite{Sale_only.2012}.
Finally in green we have the posterior distribution obtained by considering the star in isolation, i.e. without using the observations of other stars. 
Formally this is $\pr(A_i | \tilde l_i, \tilde{y}_i, \galaxymodel, \transfermodel)$.
 \label{fig:star_post}}
\end{figure}

\begin{figure}
\includegraphics{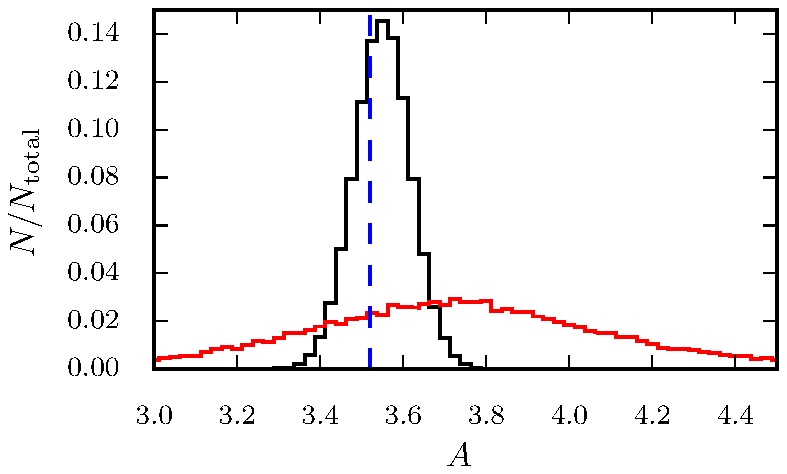}
\caption{Estimated posterior probability distributions of extinction to a position at 5000~pc and an angle of $0\arcmin$.  The `true' extinction to this point is $A_0=3.52$, as indicated by the vertical dashed blue line.
The black histogram shows the probability distribution of $\pr(A^\star|l^\star, s^{\star}, \setoflhat, \setofyhat, \micromodel, \transfermodel, \galaxymodel, \macroasumps)$ estimated using the technique described in section~\ref{sec:posterior_pred}.
The red histogram shows a comparable estimate obtained using the method of \protect\cite{Sale_only.2012}.
\label{fig:posn_post}}
\end{figure}

\section{Discussion}\label{sec:discussion}

\cite{Sale_Drew.2014} presented a 3D extinction map of the northern Galactic Plane, based on a method from \cite{Sale_only.2012}.
That method proceeds by dividing the sky into many regions approximately $10\arcmin$  in size and then estimating both mean extinction and `differential extinction' -- the unresolved variation in extinction -- as a function of distance.
One of the most striking features of the 3D extinction map it produces is that uncertainty in the extinction to a given point is dominated by the effects of unresolved substructure.
Consequently it was apparent that a method that is more directly geared towards dealing with the fractal nature of the ISM was necessary.
The method we present here fits that requirement: a physical model for the small-scale structure of the ISM is centre stage.

To demonstrate the improvements of our method with respect to \cite{Sale_only.2012}, we use the problems of estimating extinction both to particular stars and to arbitrary points in space.
We show marginal posteriors for both these problems in Figs.~\ref{fig:star_post} and~\ref{fig:posn_post}, comparing the two methods.
The difference between the two methods is best appreciated by comparing the widths of the posterior distributions: those obtained from our new method are considerably narrower and therefore more precise than those found by the older approach.
The improved precision stems directly from the more sophisticated treatment of small-scale structure in the ISM.
We also note that although the posterior expectations of extinction in both Figs.~\ref{fig:star_post} and~\ref{fig:posn_post} do not align exactly with the true values, there is no substantial bias as evidenced by Fig.~\ref{fig:5000_comp}.

A significant undesirable feature present in the maps of \cite{Marshall_Robin.2006} and \cite{Sale_Drew.2014} are `fingers of god' -- radially extended, azimuthal discontinuities in maps of dust density.
These artefacts appear because the methods of \cite{Sale_only.2012} and \cite{Marshall_Robin.2006} treat separate sight lines independently.
As our model for small scale structure of the ISM includes a spatial correlation kernel for dust density such features no longer occur.

Our decision to model the likelihood as a bivariate normal~\eqref{eq:gaussianlik} on $\{ \log A_n, \log s_n\}$ was motivated primarily by the opportunity to perform the marginalisation of $\setofA$ analytically.
The lognormal dependence of this likelihood on distance is plausibly realistic, as it corresponds to a Gaussian likelihood on distance modulus.
However, \cite{Green_Schlafly.2014} show that when we take $y_n$ to be real observables, such as a set of apparent magnitudes in different bands, then the likelihood $\pr(\tilde{y}_n | \tilde{l}_n, \tilde{b}_n, s_n, A_n, \transfermodel, \galaxymodel)$ typically takes a complicated form, reflecting the shape of stellar isochrones.
Consequently a single Gaussian in $\{\log s_n, \log A_n\}$ space is unlikely to always provide a good fit.
This can be remedied by fitting a mixture of Gaussians to each likelihood; the generalisation of our method from a single Gaussian to a mixture of Gaussians is trivial in principle and requires only some modest extra bookkeeping in practice.

The method we have described largely follows and builds upon that of \cite{Vergely_FreireFerrero.2001}.
The two methods are fundamentally similar: both employ a Gaussian random field to describe $\log A$.
However, in the models of \cite{Vergely_FreireFerrero.2001}, the large-scale ISM density is not allowed to vary.  This worked well in their study, as they only studied a relatively local volume.  Our method allows us to infer the large-scale distribution of dust, a necessary step as we seek to map extinction to much greater distances, where the distribution of dust is not well known.

Furthermore, we diverge when it comes to inferring the posterior.  We use an MCMC based approach to sample from the posterior, whilst \cite{Vergely_FreireFerrero.2001} use an iterative inversion algorithm from \cite{Tarantola_Valette.1982} to find the expectation of the posterior.
The principal benefit of the \cite{Tarantola_Valette.1982} algorithm is speed: \cite{Vergely_FreireFerrero.2001} note that they typically converge on their final solution within 10 iterations.
However, this algorithm is only valid for solving an (almost) linear least squares problem.
Consequently, the posterior, likelihood and priors can only follow Gaussian distributions.
In contrast our MCMC based algorithm is able to deal with non-Gaussian pdfs.
The method in \cite{Vergely_FreireFerrero.2001} is also unable to easily deal with uncertainties on the distance to each star, requiring that they be absorbed into extinction extinction estimates.
This will be particularly troublesome if uncertainties on a star's distance and extinction are correlated -- as is generally the case -- since it will result in a loss of precision.

Selection functions can have a particularly pathological impact on extinction mapping: a magnitude limited sample will preferentially include less extinguished stars, with the consequence that, if the selection function is not accounted for, the produced extinction map will be biased to lower extinctions \citep{Sale_only.2012}.
As selection functions contribute a strongly non-Gaussian component to the posterior, \cite{Vergely_FreireFerrero.2001} are unable to deal with them effectively.
However, they avoid this problem by only employing relatively local catalogues which are at least approximately volume limited, a small proportion of available data.
We have not directly approached selection functions in this paper:  in all the simulations we performed it was assumed that all simulated stars were `observed'.
However, we note that the method we present has the flexibility to properly deal with the selection functions which shape the catalogues we employ.  We will consider this in more detail in a future paper.

\cite{Vergely_Valette.2010} adopts a sum of two exponentials with different scales as their covariance function, whilst \cite{Lallement_Vergely.2014} employ a sum of a Gaussian and a $\mathrm{sech}$ term.
They suggest that the two scales are indicative of the ISM being shaped by two similarly strong processes which occur on different scales. 
We note that the summed kernels they employ can be used to provide a reasonable approximation of the covariance function partners of members of our $\Upsilon_{\gamma, \Omega}$ family~\eqref{eq:rhopowerspec}.
Thus we suggest that the fact they find that they need to sum two kernels arises because the simple kernels they employ provide a poor approximation of a Kolmogorov-like covariance function, but a sum of two or more simple kernels provides a reasonable approximation. 
Therefore, there is no need to invoke a second process injecting energy into the ISM.

\section{Summary and future work}\label{sec:summary}

The coming years will see vast quantities of astrophysical survey data become available from ongoing (e.g. Gaia, PAN-STARRS) and upcoming (e.g. LSST) surveys, which will join the large resources we already have from surveys such as 2MASS and SDSS.
Given the ubiquity of extinction and its corresponding imprint on the data these surveys will furnish, a precise treatment of it is vitally important in order to fully exploit these data.
Moreover, the distribution of extinction is shaped by a range of processes operating in the ISM, from the formation of spiral arms to the enrichment of the ISM by AGB stars.
Consequently the study of extinction grants us a view on these processes.

Our statistical model includes a realistic physical description of the small-scale structure of the ISM which has generally been neglected when mapping extinction.
We have assumed a fixed form for the covariance function for dust density in this paper, with the shape set by the outer scale length $\mathcal{L}_o$ and the slope of the corresponding power spectrum $\gamma$. 
However, these values are not well known and may indeed vary across the Galaxy \citep{Haverkorn_Brown.2008}. 
We intend to address this in a future paper, where we will extend the existing method to attempt to estimate these values from real observations.

We have yet to mention the wavelength dependence of reddening in this paper, or the possibility that it might vary with position.
One possible approach would be to simply assume that the form of the reddening law to a star is independent of that to all other stars.
Then it would be straightforward to marginalise over the form of the reddening law to obtain a likelihood for each star conditioned on the stars' distances and extinction, but not the form of the extinction law.
However, such an approach is clearly suboptimal in the sense that we expect that the form of the reddening law to two stars to be correlated if these two stars are located near each other.
Again we defer a deeper investigation of this issue to a subsequent paper.

We note that calculating $\pr(\setofA | \setoflhat, \setofbhat, \setofs, \macromodel, \micromodel)$ or $\pr(\tilde{\setofA} | \tilde{\setofs}, \setoflhat,\setofbhat,\setofs, \macromodel, \transfermodel, \micromodel)$, which follow multivariate lognormal distributions, is the most computationally expensive part of the MCMC algorithms employed in this paper as it involves solving for the inverse and determinant of the covariance matrix.
Moreover, naive algorithms for finding the inverse and the determinant of a matrix require $\mathcal{O}(N^3)$ time, where $N$ is the number of stars in the catalogue.
In the future we intend to apply the method we have described to large catalogues.
Clearly it will not be trivial to do so: we would have to deal with an extremely large covariance matrix, which, under a naive method would be impossible to invert on any reasonable time scale.
However, as the stars in such a catalogue will be widely spread, the overwhelming majority of pairs of stars will be at least approximately independent of each other and so the covariance matrix will be very sparse.
Therefore, it should be possible to leverage this considerable sparsity to make the method feasible on large scales.

We have demonstrated, using simulated data, that our method can successfully retrieve the distribution of extinction in a field and that it is several times more precise than the method of \cite{Sale_only.2012}.
Once the computational challenges have been overcome we will be in possession of a method for inferring three-dimensional extinction maps from large stellar catalogues,  through the use of Gaussian random fields and a physical model for the small-scale fractal structure of dust.
We will then be able to apply this method to the vast torrent of survey data to produce a beautifully precise and compelling three dimensional map of extinction.

\section*{Acknowledgements}

We would like to thank Janet Drew, Christophe Pichon, the Oxford Dynamics group and the anonymous referee for their contributions and suggestions that have improved this paper. 

We acknowledge support from the United Kingdom Science Technology and Facilities Council (STFC, ST/K00106X/1).

\bibliography{astroph_3,bibliography-2,temp}

\onecolumn
\appendix

\section{Approximating the extinction covariance function from the density power spectrum}
\label{app:covA_gen}

In this appendix we obtain the covariance function of extinction~$A$
given the power spectrum for the dust density~$\rho$.
The latter is modelled as a semi-stationary random field
\citep{Priestley_only.1965} whose power spectrum is given a
function $\upsilon(\vk)$, where this power spectrum is tapered at small~$|\vk|$.
For example, we may choose to use a member of the family $\Upsilon_{\gamma,\Omega}(\vk)$, as given by equation~\eqref{eq:rhopowerspec}.
The power spectrum is then modulated by the effects of the slowly varying functions $m_\rho(\vr)$ and
$S_\rho(\vr)$.
We use this covariance function of~$A$ to obtain an approximate expression for the covariance function of $\log A$.

We place an observer at the origin~$O$ of our coordinate system and consider the extinctions
\begin{equation}
  A(\vs_i)=\int_0^{s_i}\kappa\rho(s'_i\hat{\vs}_i)\,\d s'_i
\end{equation}
to an arbitrary pair of points $(\vs_1,\vs_2)$, in which the
$\hat{\vs}_i$ are unit vectors along the line of sight to each point
and the scalars $s_i$ measure the distances of each from the observer.
The correlation function
\begin{equation}
  \big\langle A(\vs_1)A(\vs_2)\big\rangle
  =
  \int_0^{s_1}\d s_1'\int_0^{s_2}\d s_2'
  \big\langle \kappa\rho(\vs_1')\kappa\rho(\vs_2')\big\rangle,
\end{equation}
where we have introduced $\vs'_i\equiv s'_i\hat{\vs}_i$ to keep
notation reasonably uncluttered.
As we assume that $\rho$ (strictly, $\kappa\rho$) is
semi-stationary, we may use equation~\eqref{eq:priestley} to
approximate this as
\begin{equation}
  \begin{split}
  \big\langle A(\vs_1)A(\vs_2)\big\rangle
  &\simeq
  \int_0^{s_1}\d s_1'
  \int_0^{s_2}\d s_2'
  \langle \kappa\rho(\vs_1') \rangle
  \langle \kappa\rho(\vs_2') \rangle
  \int\d^3\vk\,
    \e^{\i\vk\cdot(\vs_1'-\vs_2')}\upsilon(\vk).
\end{split}
\end{equation}
To simplify this further, we assume that the angle $\Delta\theta$
between $\hat{\vs}_1$ and $\hat{\vs}_2$ is small, and choose a
coordinate system in which $\hat{\vs}_1=(\frac12\Delta\theta,0,1)$ and
$\hat{\vs}_2=(-\frac12\Delta\theta,0,1)$.  That is, the $z$ axis of
this coordinate system bisects the lines of sight to $\vs_1$ and~$\vs_2$.
Making the approximation
$\upsilon(k_x,k_y,k_z)\simeq\upsilon(k_x,k_y,0)$ and using the Fourier-space representation of the Dirac delta,
\begin{equation}
  \delta(s)=\frac1{2\pi}\int_{-\infty}^\infty \d k_z\, \e^{\i k_zs},
\end{equation}
we obtain
\begin{equation}
  \begin{split}
    \big\langle A(\vs_1)A(\vs_2)\big\rangle
  &\simeq
2\pi
  \int_0^{\min(s_1,s_2)}\d s\,
   \langle \kappa\rho(s\hat\vs_1) \rangle
   \langle \kappa\rho(s\hat\vs_2) \rangle
  \int_{-\infty}^\infty\d k_x\int_{-\infty}^{\infty}\d k_y\,
    \e^{\frac12\i k_xs\Delta\theta}\upsilon(k_x,k_y,0).
\end{split}
\end{equation}
This can be rewritten as
\begin{equation}
  \begin{split}
  \big\langle A(\vs_1)A(\vs_2)\big\rangle
  &\simeq
2\pi
  \int_0^{\min(s_1,s_2)}\d s\,
   \langle \kappa\rho(s\hat\vs_1) \rangle
   \langle \kappa\rho(s\hat\vs_2) \rangle
   P(s\Delta\theta),
\end{split}
\label{eq:covA}
\end{equation}
where $\Delta\theta=\cos^{-1}(\hat\vs_1\cdot\hat\vs_2)$ and the function
\begin{equation}
  P(x)\equiv\int_{-\infty}^\infty\d k_x\int_{-\infty}^{\infty}\d k_y\,
    \e^{\frac12\i k_xx}\upsilon(k_x,k_y,0)
\end{equation}
can easily be tabulated for a given choice of power spectrum~$\upsilon (\vk)$.

We have made a number of assumptions to obtain this result.
First, although we have not assumed any particular form for the power
spectrum of turbulence, we do assume that it is isotropic and semi-stationary.
Second, we have made the approximation $\upsilon(k_x,k_y,k_z)\simeq\upsilon(k_x,k_y,0)$ by taking a Taylor expansion of $\upsilon(k_x,k_y,k_z)$ around $k_z=0$ and discarding all but the leading term, a valid approximation if our sightlines $s_1$, $s_2$ are much longer than~$\mathcal L_0$.
If we wished we could include further terms in the Taylor expansion, particularly if we were dealing with short sightlines.
Third, we have assumed that $\Delta\theta\ll1$, but note that satisfying the condition that $s_1,s_2\gg\mathcal L_0$ ensures that one need not worry about the small angle assumption, as $P(s\Delta\theta)\to0$ when $s\Delta\theta\gg\mathcal L_0$.

Up until now, we have made no assumptions about the Gaussianity of the
random fields $\rho$ and~$A$.  
If we now assume that $\log A$ can be modelled as a Gaussian random
field, we can follow \citet{Coles_Jones.1991} and use the result above
to approximate the covariance function of $\log A$ as
\begin{equation}
  \cov[\log A(\vs_1),\log A(\vs_2)]
  \simeq \log\left[
    \frac{\big\langle A(\vs_1)A(\vs_2)\big\rangle}
    {\big\langle A(\vs_1)\big\rangle\big\langle A(\vs_2)\big\rangle}
    \right],
\label{eq:covlogA}
\end{equation}
giving us  the relationship between the covariance function of $\log A$ and the density power spectrum.
In the following section we take the dust density model described in
section~\ref{sec:dustmodel} and test the assumption that $\log A$ can
be modelled as a Gaussian random field with the covariance function~\eqref{eq:covlogA}
derived here.

\section{Simulations of the relationship between the correlation
  functions of $\log\rho$ and $\log A$}\label{app:test}

\subsection{Simulating Gaussian Random Fields}\label{sec:simulations}

In order to test our method for mapping extinction we need to produce simulations of the ISM with realistically varying dust density. 
In order to do so we approximate $\log(\rho)$ as a GRF which we then simulate, before exponentiating to obtain a simulation of $\rho$.
This was the approach adopted by \cite{Coles_Jones.1991}, \cite{Elmegreen_only.2002} and \cite{Fischera_Dopita.2004}. 
The later two seek to simulate an absorbing screen in the ISM using particular form of a Gaussian random field: fractional Brownian motion (fBm). 

There are a number of methods for simulating Gaussian random fields, \cite{Saupe_only.1988} summarises several methods for producing fBm fields. 
A popular method, owing to its speed and ease of implementation, is the spectral synthesis method. 
This method relies on the fact that the power spectrum is the square of the Fourier transform of the field. 
As it uses Fourier transforms, this method must be used with care as this approach implicitly assumes that the field it is simulating is periodic in real space. 
However, as noted in section~\ref{sec:cov_rho}, the power spectra we employ are truncated at small $\vk$, which helps disguise the periodic nature of the simulated fields.  
Additionally, the use of Fourier transforms means that it is only possible to directly simulate stationary fields.

In the spectral synthesis method one proceeds by first taking a cube of data points in $\vk$ space. 
Each point is then assigned a complex magnitude following the square root of the power spectrum assumed and a complex phase drawn randomly in the range 0 to $2\pi$.
By then taking the Fourier transform of the cube, a data cube in real space is obtained where the data points follow a Gaussian random field with the desired power spectrum.

We would like to simulate a density field such that $|\tilde\rho(\vk)|^2 \propto \mathbf{k}^{-\gamma}$. \cite{Fischera_Dopita.2004} found that, for the range $3 \leq \gamma < 4$, a satisfactory density field can be produced by exponetiating a Gaussian random field for $\log \rho$ which exhibits a  power-law power spectrum $|\log\tilde\rho(\vk)|^2 \propto \mathbf{k}^{-\beta}$.

We can then modify the stationary dust density field by multiplying it by a slowly varying function that represents the variations in mean density, as discussed in section~\ref{sec:cov_rho}. 
\begin{figure*}
\includegraphics{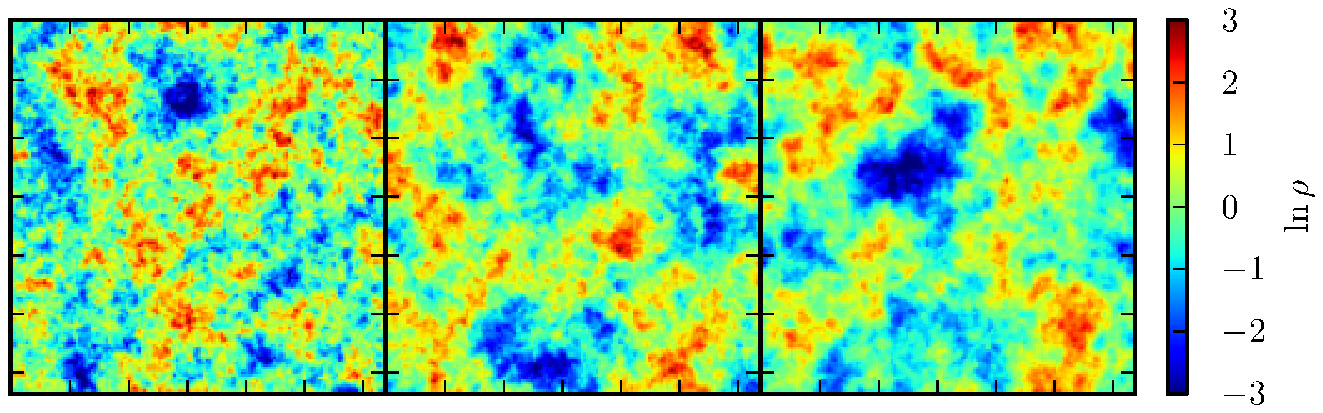}
\caption{Slices through 3 lognormal random field cubes produced by exponentiating GRFs. On the left a field simulated assuming $\gamma=3$, in the centre $\gamma=3.5$ and on the right $\gamma=4$. \label{fig:LNRF_sims}}
\end{figure*}

Following this groundwork, it is now possible to simulate dust density fields. Three example fields are shown in Fig.~\ref{fig:LNRF_sims}. Subsequently one can insert stars into the medium and integrate between the observer and the star to obtain the extinction to the star. We employ this scheme to produce the simulated data we study in section~\ref{sec:2d}.

\subsection{Comparing simulations to our assumed model}

\begin{figure*}
\includegraphics{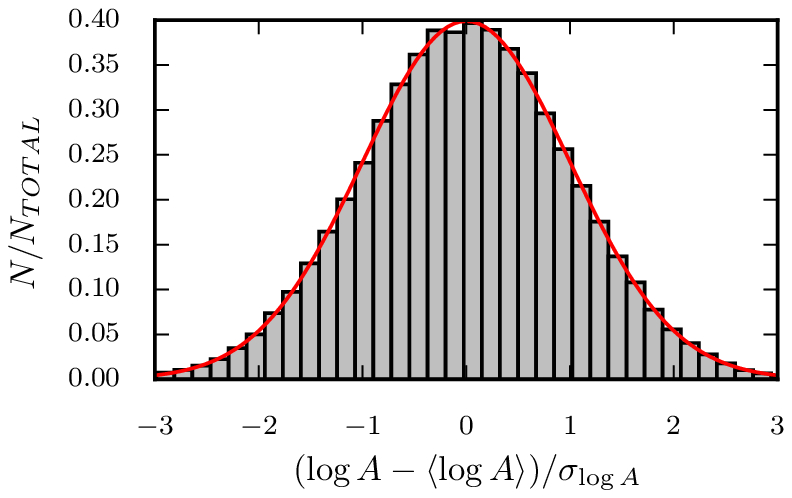}
\caption{A histogram of $\log A$ obtained at a single position for 100,000 simulations. Overplotted in red is a Gaussian distribution with a mean of zero and unit variance. \label{fig:logA_pdf} }
\end{figure*}

\begin{figure*}
\includegraphics{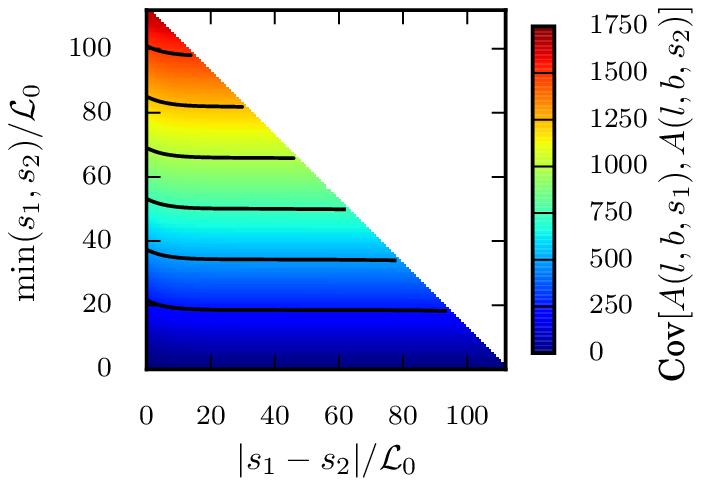}
\caption{The calculated covariance of extinction from 10,000 simulated fields. Plotted here is the dependence of covariance on the length of the shorter sightline, $\min(s_1, s_2)$,  and the difference in length between the two sightlines, $\Delta s = |s_1-s_2|$. \label{fig:min_diff_cov} }
\end{figure*}

As discussed in section~\ref{sec:dustmodel}, we assume that $\log A$ can be modelled as a Gaussian random field.
If this is true then the pdf of the logarithm of extinction to a set position in space should be Gaussian.
Here we test this assumption for the specific form of density fluctuations adopted in Appendix~\ref{sec:simulations} above, in which we assumed that $\log\rho$ was also a GRF.
To do this we simulate 100,000 dust density field using the method described above. Specifically, we assume $\gamma=11/3$ and simulate fields 2048 pixels deep and 128 wide, with a scale length of 16 pixels. We then integrate along a distance of 480 pixels, which is equivalent to 3~kpc if the outer scale length is 100~pc, to obtain a simulated extinction. Fig.~\ref{fig:logA_pdf}, shows the histogram of the simulated $\log A$, which is indeed consistent with with being Gaussian. 

The condition for a field to be a GRF is that it satsifies the condition~\eqref{eq:GRFdefn}: from this it should be clear that, in general, if $\log\rho$ satisifies this condition, it does not follow that $\log A$ does and vice versa.
However, it has been realized in a variety of contexts, such as wireless communications \citep[e.g.,][]{Mehta_Wu.2007} and finance \citep[e.g.,][]{Ju_only.2002}, that a sum of lognormal random variables can be well approximated by a single lognormal random variable.
\cite{Barakat_only.1976} describes how, due to their non-zero skew, a sum of lognormal random variables converges only slowly onto a Gaussian distribution under the central limit theorem and how, prior to convergence, the sum is better approximated by a lognormal distribution.

In the previous appendix we derived a form for the covariance function of $A$, $\cov[A(\vr_1), A(\vr_2)]$, to two positions. We showed that $\cov[A(\vr_1), A(\vr_2)]$ depends on the angular separation of the positions and the distance to the nearer position only. Consequently we have a non-stationary covariance function, where we assume that $\cov[A(\vr_1), A(\vr_2)]$ is independent of the difference between the two distances. We employ a subset of 10,000 of the 100,000 simulations produced above to study the covariance function. We consider a sub region of 1760 by 64 pixels, dropping the rest of the field to avoid the affects of periodicity in the simulations. We then calculate the covariance for every remaining pixel pair. In Fig.~\ref{fig:min_diff_cov} we show that $\cov[A(\vr_1), A(\vr_2)]$ is indeed well approximated as a linear function of the shorter distance. Moreover, we show that the covariance is sensitive to the difference in distances only in the case that the two sightlines have approximately equal length. The reason this arises is that any sections of the longer sightline that are substantially more than one scale length beyond the end of the shorter sightline will be essentially independent of the shorter sightline, whilst those sections of the longer sightline within a few scale lengths of the end of the shorter one will comprise only a small proportion of the longer sightline.

\end{document}